%% file: pengines.tex
\pdfoutput=1

\documentclass{new_tlp}
\usepackage{url}
\usepackage{html}
\usepackage{ifthen}
\usepackage{calc}
\usepackage{mathptmx}
\usepackage{color}
\usepackage{listings}
\usepackage{multirow}
\usepackage[draft]{fixme}
\usepackage{graphicx}
\newcommand{\reffont}{\tt}

\newcommand{\predref}[2]{
        \mbox{\reffont #1/#2}}

\newcommand{\secref}[1]{section~\ref{sec:#1}}
\newcommand{\Secref}[1]{Section~\ref{sec:#1}}

\newcommand{\Figref}[1]{Figure~\ref{fig:#1}}

\newcommand{\Tabref}[1]{Table~\ref{tab:#1}}

\title[Theory and Practice of Logic Programming]
        {Pengines: Web Logic Programming Made Easy}

  \author[T. Lager and J. Wielemaker]
         {TORBJ\"ORN LAGER\\
	  University of Gothenburg, Sweden \\
	  \email{Torbjorn.Lager@ling.gu.se}
	  \and JAN WIELEMAKER\\
	  VU University Amsterdam, The Netherlands\\
	  \email{J.Wielemaker@vu.nl}}

\jdate{May 2003}
\pubyear{2014}
\submitted{14 May 2014}
\revised{14 May 2014}
\accepted{27 March 2014}

\pagerange{\pageref{firstpage}--\pageref{lastpage}}

\definecolor{lightgrey}{rgb}{0.95,0.95,0.95}

\renewcommand{\arg}[1]{\textit{#1}}

\makeatletter
\def\@nodescription{false}
\newcommand{\onlinebreak}{}

\newcommand{\defentry}[1]{\definition{#1}}
\newcommand{\definition}[1]{%
	\onlinebreak%
	\ifthenelse{\equal{\@nodescription}{true}}{%
	    \def\@nodescription{false}%
	    \makebox[-\leftmargin]{\mbox{}}\makebox[\linewidth+\leftmargin-1ex][l]{\bf #1}\\}{%
	    \item[{\makebox[\linewidth+\leftmargin-1ex][l]{#1}}]}}
\newcommand{\nodescription}{\def\@nodescription{true}}
\def\predatt#1{\hfill{\it\footnotesize[#1]}}
\def\predicate{\@ifnextchar[{\@attpredicate}{\@predicate}}
\def\qpredicate{\@ifnextchar[{\@attqpredicate}{\@qpredicate}}
\def\@predicate#1#2#3{%
	\ifthenelse{\equal{#2}{0}}{%
	    \defentry{#1}}{%
	    \defentry{#1({\it #3})}}%
	\index{#1/#2}\ignorespaces}
\def\@attpredicate[#1]#2#3#4{%
	\ifthenelse{\equal{#3}{0}}{%
	    \defentry{#2\predatt{#1}}}{%
	    \defentry{#2({\it #4})\predatt{#1}}}%
	\index{#2/#3}\ignorespaces}
\def\@qpredicate#1#2#3#4{%
	\ifthenelse{\equal{#3}{0}}{%
	    \defentry{#1:#2}}{%
	    \defentry{#1:#2({\it #4})}}%
	\index{#1/#2}\ignorespaces}
\def\@attqpredicate[#1]#2#3#4#5{%
	\ifthenelse{\equal{#4}{0}}{%
	    \defentry{#2:#3\predatt{#1}}}{%
	    \defentry{#2:#3({\it #5})\predatt{#1}}}%
	\index{#2/#3}\ignorespaces}
\def\directive{\@ifnextchar[{\@attdirective}{\@directive}}
\def\@directive#1#2#3{%
	\ifthenelse{\equal{#2}{0}}{%
	    \defentry{:- #1}}{%
	    \defentry{:- #1({\it #3})}}%
	\index{#1/#2}\ignorespaces}
\def\@attdirective[#1]#2#3#4{%
	\ifthenelse{\equal{#3}{0}}{%
	    \defentry{:- #2\predatt{#1}}}{%
	    \defentry{:- #2({\it #4})\predatt{#1}}}%
	\index{#2/#3}\ignorespaces}
\newcommand{\termitem}[2]{%
	\ifthenelse{\equal{}{#2}}{%
	    \definition{#1}}{%
	    \definition{#1({\it #2})}}\ignorespaces}
\makeatother

\begin{document}

\label{firstpage}

\maketitle

  \begin{abstract}
When developing a (web) interface for a deductive database,
functionality required by the client is provided by means of HTTP
handlers that wrap the logical data access predicates. These handlers
are responsible for converting between client and server data
representations and typically include options for paginating results.
Designing the web accessible API is difficult because it is hard to
predict the exact requirements of clients. Pengines changes this
picture. The client provides a Prolog program that selects the required
data by accessing the logical API of the server. The pengine
infrastructure provides general mechanisms for converting Prolog data
and handling Prolog non-determinism. The Pengines library is small (2000
lines Prolog, 150 lines JavaScript). It greatly simplifies defining an
AJAX based client for a Prolog program and provides non-deterministic
RPC between Prolog processes as well as interaction with Prolog engines
similar to Paul Tarau's engines. Pengines are available as a standard
package for SWI-Prolog~7.\footnote{A web-based demonstration of pengines
is available at \url{http://pengines.swi-prolog.org} and can be
downloaded from \url{https://github.com/SWI-Prolog/pengines}.}
  \end{abstract}

  \begin{keywords}
web programming, query languages, agent programming, distributed
programming
  \end{keywords}


\section{Introduction}

Distributed systems play a central role in modern IT systems. We
distinguish three different models for point-to-point communication
between systems: (1) based on a query language (e.g., SQL or SPARQL),
(2) based on a generic attribute-value exchange (e.g., HTTP) and (3) based
on methods and datatypes (e.g., SunRPC, SOAP, CORBA, JSON-RPC). If
Prolog is used as a component in such systems, the typical solution is
to embed Prolog in another language through the foreign language
interface. Alternatively, Prolog may be used to implement the wire
protocol directly (often used for HTTP) or wrap a foreign library that
implements the wire protocol (e.g., ODBC).

If the task of the Prolog-based component is simple and can easily be
expressed as a deterministic function call, the above solutions are
satisfactory. If, however, Prolog is used as a (deductive) database, the
above is not ideal. In this scenario, an SQL or SPARQL like query
language over the core relational predicates provided by the database is
much more comfortable because it allows the client to specify a desired
set of results in a uniform and flexible manner. Both SQL and SPARQL
provide (1) a language to express the desired result set, (2) a way to
paginate result sets (\textit{offset} and \textit{limit}) and (3) a
uniform access to data of different types (per column in SQL and the
various RDF object types for SPARQL).

Without something similar to SQL or SPARQL, the Prolog server developer
has to imagine all sensible ways to access the data based on the
deterministic procedures and wrap these access functions in HTTP or some
form of RPC. If the application is a-priori known and has a fairly fixed
functionality, a top-down design can be satisfactory, but once this is
not the case we will typically see a large and growing set of API
functions with many options to select the proper data and represent it
in a way that is suitable for the client application. This is where
Pengines come in. The idea behind pengines is simple:

\begin{itemize}
    \item A pengine is a thread on a (often remote) Prolog pengine
	  server.
    \item The query language is Prolog, i.e., the client uploads a
	  short Prolog program to the pengine that provides the data
	  exchange needed by the client based on the clean relational
	  interface of the deductive Prolog database.
    \item Subsequently, the client sends one or more Prolog queries
          with result templates to the pengine.
    \item The pengine answers with a set of answer tuples based on
	  answer bindings of a template.  Data representation is
	  standardized.  At the moment we have two formats:
	  Prolog syntax for Prolog clients and a standard representation
	  of Prolog terms in JSON for e.g., JavaScript clients.
    \item Pagination is based on Prolog backtracking.  As an option,
          results can be batched in chunks of a certain size, e.g.,
	  return (max) 20 results per communication.
\end{itemize}

A pengine is closely related to Paul Tarau's logical engines (see
\secref{related}). If a Prolog client is used, pengines can implement
natural \textit{non-deterministic} RPC (NDRPC) as well as coroutining. The
JavaScript client allows for creating a pengine from Prolog embedded on
the HTML page, sending the pengine a query and reacting on the `answer
events'.

This article is organized as follows. First we discuss related work, in
particular the relation with logical engines as realised by Paul Tarau.
Next, we informally introduce Pengines using a number of examples. In
\secref{PLTP} we describe the state machine used to realise Pengines and
communicate with them as well as the core primitives. \Secref{derived}
describes a derived high level primitive (Prolog-RPC). Before the final
conclusions section, we address security and future plans.

\input{related.tex}

\section{Pengines by example}
\label{sec:examples}

We proceed to give a number of examples showing how pengines can be
created and controlled from any Prolog program, or from JavaScript
running in a web browser. We also show how to make non-deterministic
remote procedure calls (NDRPC) using a predicate implemented on top of
the Pengines core predicates.

\subsection{Prolog interacting with a pengine}
\label{sec:ex1}

In this example we load the Pengines library, use
\texttt{pengine\_create/1} to create a pengine in a remote
pengine server, and inject a number of clauses in it. We then  use
\texttt{pengine\_event\_loop/2}  to start an event loop that listens for
three kinds of  event terms. Running \texttt{main/0} will write  the
terms  \texttt{q(a)},  \texttt{q(b)}  and   \texttt{q(c)}  to  the
standard output of the local process.  Using \texttt{pengine\_ask/3}
with the option \texttt{template(X)}
would produce the output \texttt{a}, \texttt{b} and \texttt{c}. Removing
the \texttt{server('http://pengines.org')} option would solve the query
\texttt{q(X)} locally instead, although still concurrently.

\begin{verbatim}
:- use_module(library(pengines)).

main :-
    pengine_create(
        [ server('http://pengines.org'),
          src_text("q(X) :- p(X).
                    p(a). p(b). p(c).")
        ]),
    pengine_event_loop(handle, []).

handle(create(ID, _))           :- pengine_ask(ID, q(X), []).
handle(success(ID, [X], false)) :- writeln(X).
handle(success(ID, [X], true))  :- writeln(X), pengine_next(ID, []).
\end{verbatim}

\subsection{JavaScript interacting with a pengine}

In this example, we show  how  to   create  and  interact with a
pengine from JavaScript.   Loading the page brings up the
browser's prompt dialog, waits   for the user's input, and writes that
input in the browser  window.   If  the input was 'stop', it stops
there, else it repeats.\footnote{We could have implemented
\textit{main/0} as a recursive loop instead, but using a repeat-fail
loop nicely serves to demonstrate that Pengines gives the programmer the
same options as when programming against a simple command-line shell.}
Note that   I/O  works as expected. All we need to  do is to use
\texttt{pengine\_input/2}   instead  of \texttt{read/1}  and
\texttt{pengine\_output/1} instead of \texttt{write/1}.

\begin{verbatim}
<html lang="en">
  <head>
    <script src="/vendor/jquery/jquery-2.0.3.min.js"></script>
    <script src="/assets/js/pengines.js"></script>
    <script type="text/x-prolog">
      main :-
          repeat,
          pengine_input('myprompt>', X),
          pengine_output(X),
          X == stop.
    </script>
    <script>
      var pengine = new Pengine({
          oncreate: function() { pengine.ask('main'); },
          onprompt: function() { pengine.respond(prompt(this.data)); },
          onoutput: function() { $('#out').html(this.data); }
      });
    </script>
  </head>
  <body><div id="out"></div></body>
</html>
\end{verbatim}

\subsection{Pengines non-deterministic RPC}

Our third example shows that a non-deterministic predicate can be called
remotely by means of \texttt{pengine\_rpc/2},  yet   behave  exactly  as
if called locally:

\begin{verbatim}
?- member(X, [a, b, c, d]),
   pengine_rpc('http://pengines.org', p(X),
               [ src_list([p(b), p(c), p(d), p(e)]) ]),
   member(X, [c, d, e, f]).
X = c ;
X = d.
?-
\end{verbatim}

In this case, using the \texttt{src\_list} option, we choose to inject
four \texttt{p/1} clauses in the pengine on creation time. There are
other similar options available: the \texttt{src\_text} option will
consult a string of Prolog \textit{text} instead (see the example in
\ref{sec:ex1}), the \texttt{src\_url} option will consult the
\textit{file} referred to by a URL, and the \texttt{src\_predicate}
option will simply carry over the predicates referred to by a list of
predicate indicators to the pengine and assert them there as well.

\input{rdf.tex}

\section{The Prolog Transport Protocol}
\label{sec:PLTP}

\begin{figure}[h]
	\includegraphics[width=13cm]{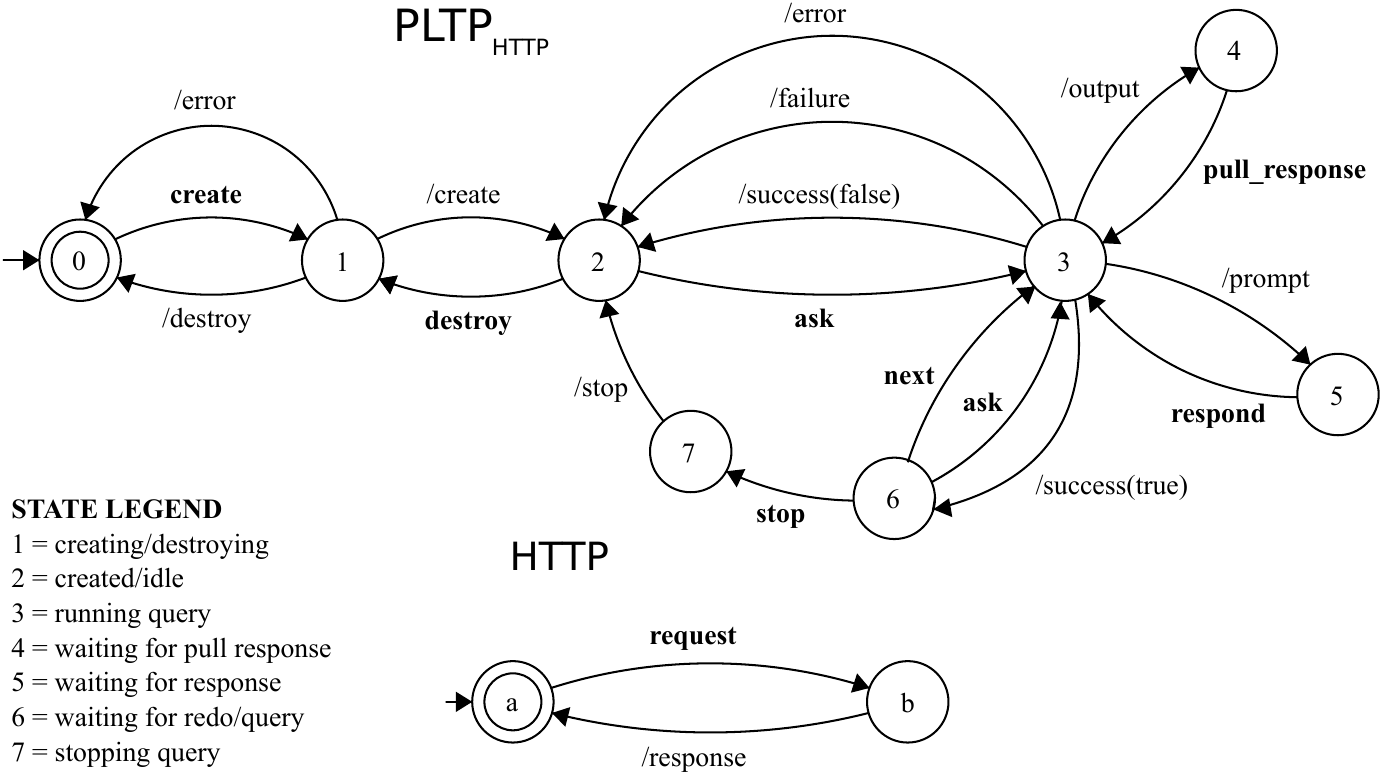}
    \caption{Communicating finite-state machines for PLTP$_{HTTP}$ and HTTP}
    \label{fig:state-machine}
\end{figure}

Underlying the design of the Pengines package is an analysis of the
conversations taking place between Prolog and a user (which could be a
human or another piece of software). Such conversations follow a
communication protocol that we refer to as the \textit{Prolog Transport
Protocol} (PLTP). The protocol is based on the Prolog 4-port model,
\cite{byrd:80} extended with exceptions and I/O and has been modelled by
means of so called \emph{communicating finite-state machines}
\cite{brand1983communicating}. A slight modification of the protocol,
referred to as PLTP$_{HTTP}$, makes it compatible with HTTP, where only
the client can take initiative. \Figref{state-machine} depicts the
communicating finite-state machines for PLTP$_{HTTP}$ and HTTP. Labels
in bold indicate requests, and labels with a slash in front indicate
responses.

\begin{figure}[h]
	\includegraphics[width=13cm]{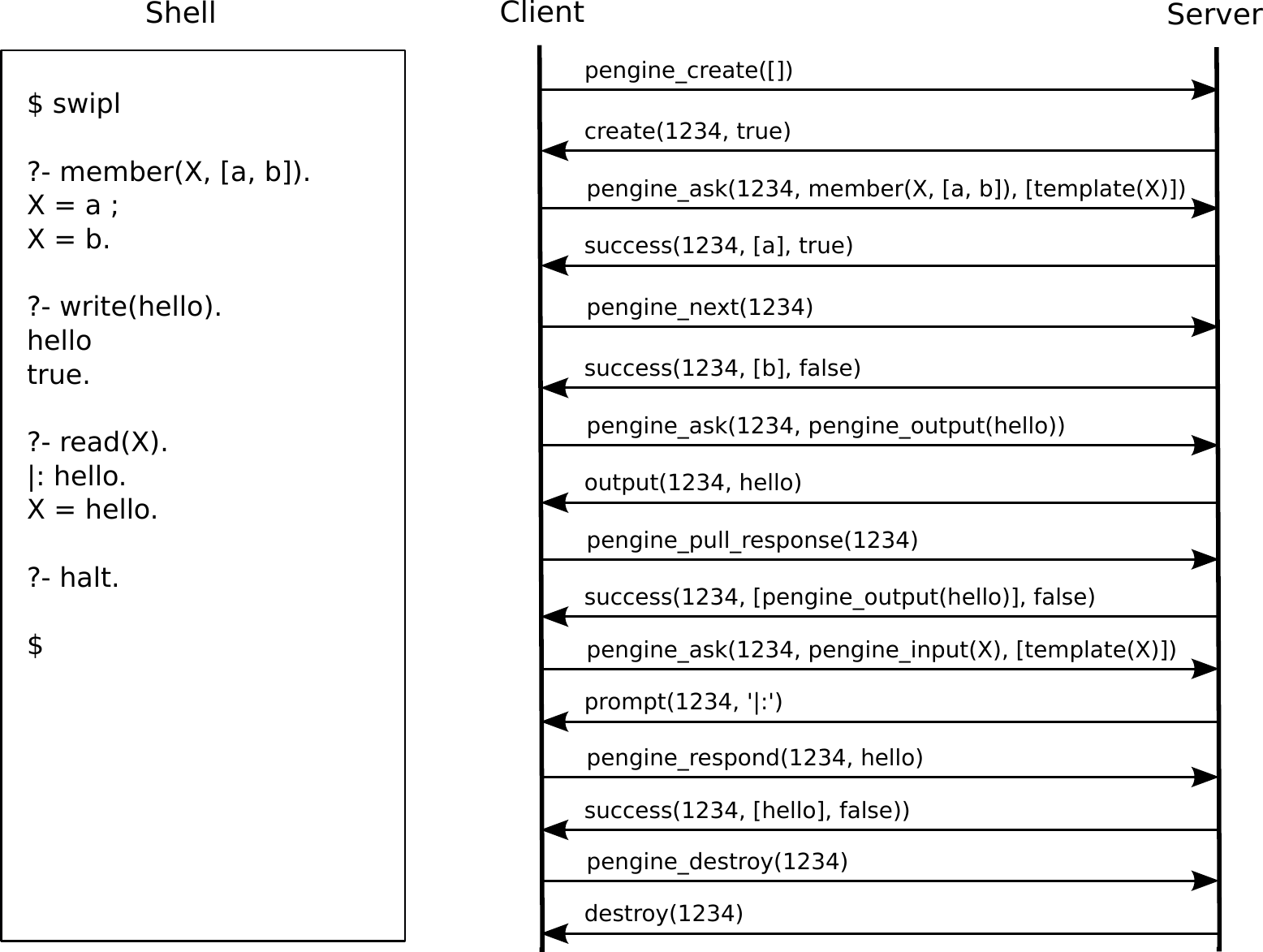}
    \caption{The PLTP run (on the right) corresponding to a user's
    interaction with Prolog (on the left). `1234' is the Pengine's
    identifier, which is a UUID in the actual implementation.}
    \label{fig:run}
\end{figure}

Figure \ref{fig:run} shows the traffic between a client and a server on
which a pengine (with ID 1234) is first created, then used to solve
queries involving both backtracking and I/O, and finally destroyed.

\section{The Prolog client: Core predicates}

We distinguish the pengine \textit{core} predicates from
\textit{derived} predicates defined in terms of the core. The core
predicates are listed in the left column of Table \ref{pl-table}. We
deal with some of the derived predicates in Section \ref{sec:derived}.

Each core predicate corresponds to a
request in the PLTP protocol (see the transitions with labels
set in bold in Figure \ref{fig:state-machine}). The terms listed in the
right column of Table \ref{pl-table} are returned as responses to such
requests.

\begin{table}[h]
  \caption{Prolog requests and responses.  Requests result in a single
	   response, but the response depends on the program behaviour.
	   \Figref{state-machine} defines which responses are possible
	   for each request.}
  \label{pl-table}
  \begin{minipage}{\textwidth}
    \begin{tabular}{ll}
      \hline\hline
      Requests& Responses\\
      \hline
      \texttt{pengine\_create(Options)}	      &	\texttt{create(ID,Data)}       \\
      \texttt{pengine\_ask(ID,Query,Options)} &	\texttt{output(ID,Data})       \\
      \texttt{pengine\_pull\_response(ID)}    &	\texttt{prompt(ID,Data)}       \\
      \texttt{pengine\_respond(ID,Term)}      &	\texttt{success(ID,Data,More)} \\
      \texttt{pengine\_next(ID)}	      &	\texttt{failure(ID)}	       \\
      \texttt{pengine\_stop(ID)}	      &	\texttt{error(ID,Data)}	       \\
      \texttt{pengine\_destroy(ID)}	      &	\texttt{stop(ID)}	       \\
      \texttt{pengine\_abort(ID)}	      &	\texttt{destroy(ID)}	       \\
      \hline\hline
    \end{tabular}
    \vspace{-2\baselineskip}
  \end{minipage}
\end{table}

Several pengine core predicates accept options that modify or control the
creation or run time behaviour of pengines. We describe those that we
believe are the most important ones. For \texttt{pengine\_create/1}
there is an option \texttt{id(-ID)} that will bind \texttt{ID} to a
UUID that can be passed around and
that allows any process that knows it to enter into a conversation with
the pengine. The \texttt{name(+Atom)} option allows a master process to
name its slave pengines, and to use simple user defined names instead of
opaque system generated identifiers in future conversations. When a
process creates a pengine remotely, with a request over HTTP, it
receives the identifier in the response. In order to save network
round-trips, Prolog code in the form of a list of clauses, Prolog text,
or a URL resolving to a \texttt{*.pl} file accessible on the Web can be
injected into the pengine at creation time.

The \texttt{pengine\_ask/3} predicate also accepts options. The
\texttt{template(+Template)} option allows the invoker of a query to
select the interesting parts of solutions through variables shared with
the query. A list of instantiations of \texttt{Template} (rather than a
list of instantiations of the whole query) will appear in the second
argument of the \texttt{success} event terms returned to the invoker by
the pengine. The \texttt{chunk(+N)} option allows the invoker to
retrieve solutions in chunks of \texttt{N} from the pengine, rather than
the default one by one. As we shall see in \secref{efficiency}, this
mechanism can be used not only to paginate the solutions generated by a
query to (say) a deductive database, but also to optimise performance by
allowing  to make fewer network round-trips.

\section{The JavaScript client}

The JavaScript client code consists of a definition of a single
\texttt{Pengine} object, written in less than 150 lines of JavaScript.
When an object is created, a pengine is (by default) created on the
pengine server from which this client code was loaded. The object
exposes a number of methods that allows the client to send requests
(queries, input, commands, etc.) to the pengine. By default, pengine
responses reach the client in the form of JSON events to be handled by
JavaScript functions. Table \ref{js-table} lists the main methods
exposed by the JavaScript object, as well as the form of JSON responses.

\begin{table}[h]
  \caption{JavaScript requests and responses}
  \label{js-table}
  \begin{minipage}{\textwidth}
    \begin{tabular}{ll}
      \hline\hline
      Requests& Responses\\
      \hline
      \texttt{pengine = new Pengine(\textit{options})}& \texttt{\{"event":"create", "id":\textit{ID}, "data":\textit{Data}\}}              \\
      \texttt{pengine.ask(\textit{query}, \textit{options})}   & \texttt{\{"event":"output", "id":\textit{ID}, "data":\textit{Data}\}}              \\
      \texttt{pengine.pull\_response(\textit{ID})}    & \texttt{\{"event":"prompt", "id":\textit{ID}, "data":\textit{Data}\}}              \\
      \texttt{pengine.respond(\textit{term})}           & \texttt{\{"event":"success", "id":\textit{ID}, "data":\textit{Data}, "more":\textit{Bool}\}}\\
      \texttt{pengine.next()}                & \texttt{\{"event":"failure", "id":\textit{ID}\}}                          \\
      \texttt{pengine.stop()}                & \texttt{\{"event":"error", "id":\textit{ID}, "data":\textit{Data}\}}               \\
      \texttt{pengine.destroy()}             & \texttt{\{"event":"stop", "id":\textit{ID}\}}                             \\
      \texttt{pengine.abort()}               & \texttt{\{"event":"destroy", "id":\textit{ID}\}}                          \\
      \hline\hline
    \end{tabular}
    \vspace{-2\baselineskip}
  \end{minipage}
\end{table}

For data, Pengines defines the following mapping between  ground Prolog terms and
JSON:\footnote{This mapping will be reconsidered with the introduction
of strings and dicts in SWI-Prolog version~7, see \url{http://www.swi-prolog.org/pldoc/man?section=extensions}.}

\begin{enumerate}[(iii).]
    \renewcommand{\theenumi}{(\roman{enumi})}
    \item A Prolog atom is mapped to a JSON string.
    \item A Prolog number is mapped to a JSON number.
    \item A Prolog list is mapped to a JSON array.
    \item The Prolog terms \texttt{@(true)}, \texttt{@(false)} and \texttt{@(null)} are mapped to the
          JSON constants \texttt{true}, \texttt{false} and \texttt{null}, respectively.
    \item A Prolog term \texttt{json(NameValueList)}, where \texttt{NameValueList} is a
          list of \texttt{Name=Value} pairs, is mapped to a JSON object.
    \item Any other complex Prolog term \texttt{T} is mapped to a JSON object of
          the form \texttt{\{"functor": F, "args": A\}} where \texttt{F} is a string
          representing the functor of \texttt{T} and \texttt{A} is the list of JSON values
          representing \texttt{T}'s arguments.
\end{enumerate}

\noindent For example, the JSON response encoding the second solution to
\texttt{append(Xs,Ys,[a,b,c])} is:

\begin{verbatim}
 { "event":"success", "id":1234, "more": true,
   "data":[{"Xs":["a"], "Ys":["b","c"]}]
 }
\end{verbatim}

\section{The Prolog client: Derived predicates}
\label{sec:derived}

The core pengine predicates are   deterministic, yet they can control
one or more pengines solving possibly non-deterministic queries. The
package also offers a number of derived  non-deterministic predicates,
built on top of the deterministic ones, that can  solve queries "the
Prolog way", binding query variables in the process, backtracking for
more solutions. Of these predicates, \texttt{pengine\_rpc/3} is the
most important. By means of \texttt{pengine\_rpc/3} a pengine running in
a pengine server A can call and try to solve a query in the  context
of  another pengine server B, taking advantage of the data being offered
by B, just as if the data was local to A.

Implementing \texttt{pengine\_rpc/3} involves creating a remote pengine,
which may send additional predicates to the remote server and
subsequently realise the normal declarative and operational semantics of
running a Prolog predicate by acting on the pengine events.

\section{Efficiency considerations}
\label{sec:efficiency}

Efficiency considerations when programming with pengines are important.
There are two major sources of inefficiency related to network
programming: network latency and bandwidth. Bandwidth requirements are
limited by bringing the code to the data. E.g., consider solving
\verb$p(X,Y),q(Y,Z)$ where \arg{Y} is a large intermediate result that
is not needed by the client. By default, Pengines will transport all
variable bindings and thus also \arg{Y}. This can be avoided in one of
two ways. Either we can define a predicate \texttt{pq(X,Z)}, or we
can use the option \texttt{template(Z)} to only transport the binding
for \arg{Z}.

The impact of network latency is limited by reducing the number of round
trips. This is achieved with three features. First of all, the first
query may be passed with the \textit{create} command. Second, the
pengine by default self-destroys itself on determinstic completion,
failure or error of the initial query. Third, the \texttt{chunk(N)}
option, sends answers in chunks of the specified size. Calling
\texttt{pengine\_rpc(URL, Query, [chunk(N)])} will result in a call to
\texttt{find\_n(N, Query, Query,
Solutions)}\footnote{\texttt{find\_n(+N, ?Template, +Goal, ?List)} acts
like \texttt{findall/3} but returns only the first \texttt{N}
bindings of \texttt{Template} to \texttt{List}, on backtracking another
batch of \texttt{N} bindings, and so on.} on the pengine server located
at \texttt{URL}.

\section{Security}
\label{sec:security}

There are three layers of safety that relate to pengines. First of all, we
can rely on the safety of the underlying HTTP protocol. Considering that
Prolog has full access to the OS, this is like using an outdated
unencrypted telnet session to the client. Alternatively, we can use
HTTPS with authentication, which makes it similar to SSH access to a
shell.

To facilitate anonymous users, SWI-Prolog provides
library(sandbox),\footnote{\url{http://www.swi-prolog.org/pldoc/doc/swi/library/sandbox.pl}}
which exports \predref{safe\_goal}{1}. This predicate performs abstract
interpretation of the argument goal and either succeeds, throws an
\textit{instantiation\_error} if it cannot prove sufficient
instantiation to a meta-goal or throws a \textit{permission\_error} if
it encounters a possibility that a predicate may be called that is not
white-listed. The library provides a predefined whitelist consisting of
pure Prolog built-in predicates, which enables it to prove the safety of
many of the pure Prolog libraries.

The above is typically sufficient if the server uses plain Prolog. If it
uses, e.g., the SWI-Prolog RDF database, the C-defined query facilities
of this library can be added to the whitelist.

Finally, by means of various application settings the Pengines library also offers some protection against denial of service (DoS) attacks. One setting determines the number of pengines that can be run simultaneosly by an application, another setting the maximum number of slave pengines that a master process may create, and a third setting is a timeout that aborts the computation after a set time, thus protecting against runaway computations.

\section{Portability and interoperability}
\label{sec:portability}

The current implementation of pengines is heavily based on support for
multiple threads as well as the SWI-Prolog HTTP server and client
libraries \cite{wielemaker:tplp2008}. It can probably be ported fairly
easily to YAP. Porting to other Prolog implementations that implement
the ISO working draft for
threads\footnote{\url{http://logtalk.org/plstd/threads.pdf}} is probably
feasible if the systems provide HTTP server and client access. It is also
possible to realise the protocol using Prolog processes managed from a
conventional HTTP server. Such a coarse grain implementation is more
robust against malicious pengines, but individual pengines start much
slower and cannot easily share a Prolog database.

To achieve maximal interoperability between pengine platforms, that
will allow a client on one platform to send and execute code on another
platform, an effort must be made to standardise a subset of Prolog that
runs in the same way on all platforms. Fortunately, such an effort only
needs to deal with a safe subset of Prolog (see Section
\ref{sec:security}).

\section{Evaluation}
\label{sec:evaluation}

In this section we present insight into the overhead involved in using
pengines. \Tabref{basictiming} shows the basic HTTP overhead, a minimal
pengine RPC call and the execution time of the example from \secref{rdf}
in three scenarios. We minimised network overhead by using connections
to \textit{localhost}. Timings were performed on a machine with an Intel
Core i7-3770 CPU running Ubuntu 13.10 (kernel 3.11) and SWI-Prolog
7.1.14. Times are in milliseconds, taking the average of 10 runs, each
with 1,000 iterations.

\begin{table}[h]
    \caption{Basic timing}
    \label{tab:basictiming}
\begin{tabular}{p{0.6\linewidth}rr}
\hline\hline
Test	& CPU time (ms) & Wall time (ms) \\
\hline
Most simple HTTP request                      & 0.4 & 0.8 \\
RPC for \texttt{true}                         & 0.9 & 1.9 \\
\predref{event\_in\_area}{3} on server        & 7     &	7 \\
RPC \predref{event\_in\_area}{3}, chunk = 1   & 159   & 386 \\
RPC \predref{event\_in\_area}{3}, chunk = 128 & 11    & 39 \\
\hline\hline
\end{tabular}

\end{table}

\begin{figure}[h]
	\includegraphics[width=0.8\linewidth]{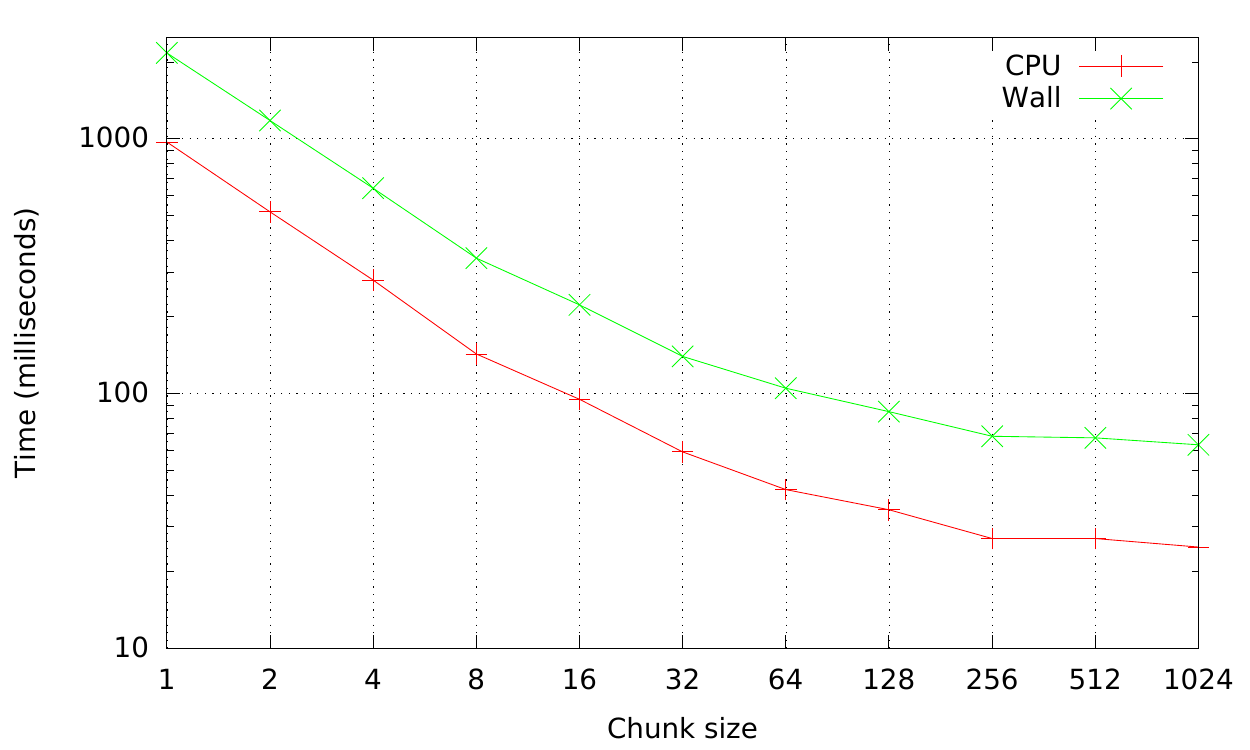}
    \caption{Performance with varying chunk sizes}
    \label{fig:chunking}
\end{figure}

The effect of chunking the result set is significant because fetching
the next result batch involves an HTTP request.  Nevertheless, using a
high value may not be the optimal choice if not all answers are needed.
\Figref{chunking} illustrates the elapsed and CPU time used for
fetching all 1981 solutions for \predref{event\_point}{2} with different
chunk sizes.

\section{Future plans}
\label{sec:plans}

The concept of pengines is simple and is not likely to change much. One
addition that we do however contemplate is a way to allow two
pengines that are created and ``owned" by different masters to
communicate. Currently, this is not possible.

Two areas need further attention: security and performance. The
current security model aims at retrieving data from the server without
compromising it. Future versions are likely to include support to deny
access to certain data and provide authorized access to certain data,
which may include update operations. Regarding performance, it is easy
to get into scenarios where the HTTP protocol and network connection
overhead become severely limiting factors, in particular for
\predref{pengine\_rpc}{3}. Protocol and connection overhead can be
reduced by using WebSockets.\footnote{\url{http://www.websocket.org/}}
We also plan to implement \emph{dynamic} chunking of the results. This
could either be the \predref{pengine\_rpc}{3} client dynamically
switching to larger chunks or the server computing additional answers
while waiting for a \textit{next} command and returning all available
answers when the \textit{next} arrives.

\section{Conclusions}

Pengines extend Tarau's Prolog
engines by allowing for Prolog engines to live in a remote (Prolog)
process. The communication with pengines is based on a state machine and
uses either Prolog or JSON for serialization of messages. The current
transport protocol is HTTP, which facilitates web applications and
allows pengines to communicate smoothly with common firewall
configurations. Prolog interaction with pengines can be used to
implement agent interaction protocols, control structures such as
coroutining blocks as well as convenient non-deterministic Prolog RPC.
JSON based interaction constitutes an ideal way to interface Prolog with
JavaScript for implementing web applications.

Creating a remote pengine may involve uploading a Prolog program to the
server. `Bringing the code to the data' greatly reduces data exchange
between client and server. It allows the server to implement only a
clean relational interface, while the client can ask questions that
involve multiple relations efficiently.

Pengines are versatile. Applications include providing a web interface
for a classical question/response Prolog application, complex network
transparent control structures, Prolog-to-Prolog RPC, a modular
alternative to the SPARQL query language for the semantic web or a web
application based on a Prolog server.

We would like to stress that Pengines is not just yet another library
for SWI-Prolog. We believe our work can be viewed more abstractly, as a
description of a general approach to web logic programming, that can be
given a concrete manisfestation not only for Prolog but also for other
one-tuple-at-a-time logic programming languages.

\bibliographystyle{acmtrans}
\bibliography{pengines}

\label{lastpage}



\end{document}

%% file: related.tex
\section{Related work}
\label{sec:related}

The notion of explicit Prolog engines has been explored extensively by
Paul Tarau et al \cite{tarau2009interoperating} in the context of the
Jinni agent programming language and the BinProlog system. Tarau's
engines are in-process and primarily designed to provide a clean
alternative implementation for Prolog language constructs such as the
all-solutions predicates (e.g., findall/3), exception handling as well
as language constructs that are less common in the Prolog world, such as
multiple coroutining blocks.

In contrast, \textit{pengines} are designed primarily for creating and
accessing Prolog engines on a remote server and communicating with them
using multiple languages (currently Prolog and JavaScript).  Tarau's
engine API and the Prolog implementation of our Pengine client are
closely related.  We explain the differences below.

\begin{description}
    \item[Creation]
Tarau's engines are created using a \arg{Goal} and \arg{AnswerTemplate}.
As our pengines typically run remote, they are more heavy weight and we
decided that a created pengine can be used to execute multiple queries
using \predref{pengine\_ask}{3}, i.e., a pengine is an engine that
runs a goal that asks for goals to execute. Typically, a pengine is
created with a \textit{Prolog source}, providing the code to execute.
This is irrelevant for Tarau's engines as they run in the same process.
(Remote) pengine creation requires some additional information, such as
the address of the server and the data-format for exchanging events
(Prolog or JSON).

    \item[Yield/return]
Tarau's yield and return primitives are represented using
\predref{pengine\_output}{1} (return) and \predref{pengine\_input}{2}
(yield).

    \item[One-sided communication]
Designed to run over HTTP, only the `client' can take initiative in the
pengines world. The \predref{pengine\_pull\_response}{2} primitive
realises bi-directional initiative, based on the `long polling'
technique that is commonly used in the context of HTTP to achieve server
initiative. This is not needed if engines
are embedded in the same process or can use a bi-directional
communication channel. (See also \secref{plans}.)
\end{description}

Pengines can also be regarded as a high-level interface to Prolog,
similar to InterProlog \cite{DBLP:conf/jelia/Calejo04} and the
multi-language interface supported by ECLIPSE Prolog
\cite{DBLP:conf/padl/ShenSNS02}. Whereas InterProlog provides a
Java-Prolog interface, ECLIPSE supports Tcl/Tk and Visual Basic as well
as Java. At this time, Pengines only supports Prolog-Prolog and
JavaScript-Prolog communication, the latter of utmost importance for a
seamless integration between Prolog and the Web. Other languages can
easily be added. Pengines also handles backtracking over
non-deterministic queries as well as I/O, something which is not
supported by the other interfaces.

%

%% file: rdf.tex
\subsection{Using pengines as a semantic web query language}
\label{sec:rdf}

Our last example is a realistic example that illustrates how pengines
can be used to make deductive databases available over the web. We build
this example on top of an existing demo application for the semantic
web. The core of the semantic web is the data description language
RDF.\footnote{\url{http://www.w3.org/TR/2004/REC-rdf-primer-20040210/}}
An RDF model consists of a set of triples of the form
\{\arg{Subject}, \arg{Predicate}, \arg{Object}\}. The ClioPatria
system\footnote{\url{http://cliopatria.swi-prolog.org}} provides a
storage layer for RDF that makes the RDF triples available using the
declarative relation \predref{rdf}{3}.

This example illustrates the following benefits from pengines: (1)
adding pengines provides a generic API to the application with only a
few lines of code and (2) available relations on the server can be
combined using Prolog control structures that are executed on the
server.

The short code fragment below adds pengine support to ClioPatria. After
loading the pengines service, we load the RDF libraries into the pengine
context to make them accessible. Next, we need to tell the sandboxing
library that the predicate \predref{rdf}{3} is \emph{safe}, i.e.,
executing it does not pose a security risk.

\begin{verbatim}
:- use_module(library(pengines)).
:- use_module(pengine_sandbox:library(semweb/rdf_db)).
:- use_module(pengine_sandbox:library(semweb/rdfs)).
:- use_module(library(sandbox)).

:- multifile sandbox:safe_primitive/1.
sandbox:safe_primitive(rdf_db:rdf(_,_,_)).
\end{verbatim}

For the remainder of this example, we populate ClioPatria with the `Open
Piracy'
data\footnote{\url{http://cliopatria.swi-prolog.org/help/source/doc/home/vnc/prolog/src/ClioPatria/web/tutorial/Piracy.txt}}.
The Open Piracy dataset contains information about piracy events around
the world. The central notion is an `event' that has a type, involved
actors, a place and a time. We will use \predref{pengine\_rpc}{3} to
query the ClioPatria service about events. First, we define simple
building blocks for querying the dataset, that (1) relate an event to
its position on the globe and (2) test that the event takes place in a
given bounding box on the globe. Note that the code below could be
loaded a priori into the ClioPatria instance providing a high-level
domain specific query language, or can be uploaded from the client.

\begin{verbatim}
event_point(Event, point(Lat, Lon)) :-
    rdfs_individual_of(Event, sem:'Event'),
    rdf(Event, sem:hasPlace, Place),
    rdf(Place, wgs84:lat, literal(type(xsd:decimal, LatText))),
    rdf(Place, wgs84:long, literal(type(xsd:decimal, LonText))),
    atom_number(LatText, Lat),
    atom_number(LonText, Lon).

event_in_area(Event, point(Lat,Lon), area(LatMin,LonMin,LatMax,LonMax)) :-
    event_point(Event, point(Lat, Lon)),
    Lat >= LatMin, Lat =< LatMax,
    Lon >= LonMin, Lon =< LonMax.
\end{verbatim}

Given the code above, the client can query for events in a given
bounding box using the code below. The \texttt{src\_predicates}
option transfers the given predicates from the client to the pengine on
the server.

\begin{verbatim}
?- pengine_rpc(
       'http://openpirates.org',
       event_in_area(Ev, Point, area(2,45,20,55)), [
           src_predicates([event_point/2, event_in_area/3])
       ]).
\end{verbatim}

Note that the execution of the \predref{event\_in\_area}{3} predicate
takes place on the server.  The execution consists of a conjunction of
several \predref{rdf}{3} goals, data transformation and conditions, yet
only results are passed back to the client.